\documentclass[11pt]{article} 

\usepackage[margin=1in]{geometry}
\usepackage{setspace}

\normalsize
\usepackage{amsfonts}
\usepackage{amssymb}
\usepackage[cmex10]{mathtools}
\usepackage{amscd}
\usepackage{graphics}
\usepackage[dvips]{graphicx}
\usepackage{epsfig}
\usepackage{array}
\usepackage{eqparbox}
\usepackage{hyperref}
\usepackage{fancyhdr}
\usepackage{appendix}
\usepackage{titling}
\usepackage{bm}
\usepackage{enumerate}
\usepackage{tabularx}
\usepackage[super]{nth}
\usepackage{caption} 
\usepackage{subcaption}

\usepackage{booktabs}
\usepackage{float} 

\usepackage[round]{natbib}  
\DeclareMathAlphabet{\mathbsf}{OT1}{cmss}{bx}{n}
\DeclareMathAlphabet{\mathssf}{OT1}{cmss}{m}{sl}
\DeclareMathAlphabet{\mathcsf}{OT1}{cmss}{sbc}{n}

\usepackage{stackengine}
\newcommand{\ie}{{\em i.e.}}

\newcommand{\secref}[1]{Section~\ref{#1}}

\newcommand{\tabref}[1]{Table~\ref{#1}}
\newcommand{\thrmref}[1]{Theorem~\ref{#1}}

\newcommand{\keywords}[1]{\textbf{Keywords:} #1}

\makeatletter
\def\blfootnote{\xdef\@thefnmark{}\@footnotetext}
\makeatother

\newtheorem{theorem}{Theorem}[section]

\newcommand{\qed}{\nobreak \ifvmode \relax \else
      \ifdim\lastskip<1.5em \hskip-\lastskip
      \hskip1.5em plus0em minus0.5em \fi \nobreak
      \vrule height0.75em width0.5em depth0.25em\fi}

\usepackage{newlfont}
\date{}
\begin{document}
\title{\Large{\textbf{Statistical two-round search for one excellent element}}}
\author{Nagananda K G and Jong Sung Kim\thanks{The authors are with Fariborz Maseeh Department of Mathematics and Statistics, Portland State University, Portland, OR 97201, USA. E-mail: \texttt{\{nanda, jong\}@pdx.edu}.}}
\setlength{\droptitle}{-1.in}
\maketitle
\vspace{-2cm}

\begin{abstract}
We formulate and study a statistical version of Katona's two-round search problem of finding at least one excellent element in a set.  A population of $n$ elements is considered, where each element is independently excellent with probability $\lambda/n$, $\lambda > 0$. A subset test is noiseless: it returns positive exactly when the queried subset contains at least one excellent element.  The goal is to minimize the expected number of tests subject to finding one excellent element with probability at least $1-\alpha$, where $0<\alpha<1$, under the restriction that testing is performed in two rounds.  Unlike classical group testing, the objective is not to recover the full set of excellent elements, but only to identify one of them.  We first show that success is fundamentally limited by the possibility that no excellent element exists.  In the sparse Poisson regime, this imposes the necessary feasibility condition $\alpha\ge e^{-\lambda}$.  When the target success probability is feasible, we prove that the optimal expected number of tests grows logarithmically with the population size.  The upper bound is obtained by combining an initial existence test with a second-round separating design; the lower bound follows from an information-counting argument.  Numerical illustrations show the feasibility boundary and the resulting logarithmic scaling.
\end{abstract}

\vspace{-0.25cm}
\section{Introduction}\label{sec:introduction}\vspace{-0.3cm}
The problem of finding an ``excellent'' element from a finite population is a basic search problem with a distinctive combinatorial structure; see, for instance, the work of \cite{Katona1973}, \cite{Ahlswede1987}, \cite{Aigner1988}, \cite{Du2000}, \cite{Gerbner2018}, \cite{Damaschke2019}.  Within this class of problems, \cite{Katona2011} proposed a two-round formulation in which one is given a set of $n$ elements containing an unknown number of excellent elements at unknown positions.  The task is to identify at least one excellent element using subset queries. A query consists of selecting a subset of elements and asking whether that subset contains an excellent element.  The search is carried out in two stages. In the first stage, the tester must choose a collection of subsets without seeing any test results. The answers to these first-stage tests are then revealed. Using only this information, the tester chooses a second collection of subsets for the final stage. After the second-stage answers are revealed, the tester must identify at least one excellent element, if the available information permits. The main question is how to choose the two collections of subsets so that at least one excellent element can be found using as few tests as possible.  A variant of this problem was investigated by \cite{Gupta2026} wherein at most one of the answers of the tests might be incorrect\textemdash it is referred to as a ``lie''; also see the work of \cite{Renyi1961}, \cite{Ulam1976} in this setting. 

In this paper, we formulate and study a statistical version of the two-round problem posed by \cite{Katona2011}. Instead of assuming deterministically that excellent elements exist (or, do not exist) in a set, we assume that excellence is random and rare. Specifically, given a population of $n$ elements, each element is independently excellent with probability $p_n = \lambda/n$, and $\lambda>0$.  Thus the expected number of excellent elements is $\lambda$, independent of $n$. This is the natural sparse asymptotic regime: excellent elements are rare, but the total number of excellent elements remains of constant order as the population size grows.  In this statistical setting, the event that the population contains no excellent element has positive probability and is explicitly part of the model. Consequently, the goal cannot be unconditional identification of an excellent element. Instead, our aim is to design a two-round testing procedure that succeeds with prescribed probability at least $1-\alpha$, for $0 < \alpha < 1$, while minimizing the expected number of subset tests.

A statistical motivation arises from large-scale screening and experimental design problems in which the objective is discovery rather than full classification. In many planned investigations, the investigator does not need to identify all active or superior units; it may be sufficient to find one unit that meets a pre-specified excellence criterion. Examples include selecting one promising treatment combination from a large factorial or combinatorial library explained by \cite{Ghose2001}, identifying one active component in a high-throughput screening experiment, or locating one anomalous unit in a quality-control population. In such problems, excellent elements are typically rare, and a sparse Bernoulli model provides a tractable statistical description of this rarity. Subset tests may be interpreted as pooled experiments or aggregate assays, where a positive response indicates that at least one member of the tested subset is excellent. The two-round restriction reflects a common planning constraint: experiments are often run in batches, with the second batch designed only after the first-stage results are observed. Thus the problem combines statistical modeling, constrained experimental planning, and optimal test allocation. The objective is not full recovery of the excellent set\textemdash as in classical group testing\textemdash but the more targeted goal of finding one excellent element with prescribed probability while minimizing expected testing effort.

\vspace{-0.25cm}
\subsection{Problem formulatiom}\label{subsec:problem_formulation}
Let $[n] = \{1,2,\ldots,n\}$ denote the population of elements. For each $i \in [n]$, let
\begin{equation*}
X_i = 
\begin{cases} 
1, & \text{if element } i \text{ is excellent}, \\ 
0, & \text{otherwise}. 
\end{cases}
\end{equation*}
We assume $X_1,\dots,X_n$ are independent Bernoulli random variables with common success probability $p_n = \lambda/n$, with $\lambda>0$ fixed. Let $E_n = \{i\in[n]:X_i=1\}$ be the random set of excellent elements. A subset test applied to a set $A \subseteq [n]$ returns the noiseless binary response $Y(A) = \mathbf{1}\left\{A\cap E_n \neq \varnothing\right\}$.  Thus a test is positive if and only if the tested subset contains at least one excellent element.

A two-round statistical search procedure consists of the following.  First, before observing any test outcomes, the procedure selects a collection of first-round test sets $\mathcal{A}_1 = \{A_{1,1},\ldots,A_{1,m_1}\}$,  $A_{1,j} \subseteq [n]$; $j = 1, \dots, m_1$. The test sets are allowed to have arbitrary sizes. Thus the formulation does not require $\lvert A_{1,1}\rvert, \ldots, \lvert A_{1,m_1} \rvert$ to be equal. The subsets may also overlap. This generality is intentional: it permits arbitrary two-round subset-testing designs. Special cases such as partitions of $[n]$ or test families with equal block sizes may be considered as particular designs, but they are not imposed in the optimization problem.  After observing the first-round response vector $Y_1 = \left(Y(A_{1,1}), \dots, Y(A_{1,m_1})\right)$, the procedure selects a second-round collection of test sets $\mathcal{A}_2(Y_1) = \{A_{2,1}(Y_1),\ldots,A_{2,m_2(Y_1)}(Y_1)\}$,  $A_{2,k}(Y_1)\subseteq[n]$; $k = 1, \dots, m_2$.  The number of second-round tests is allowed to depend on the first-round outcomes. After observing the second-round responses, the procedure outputs either an element $\widehat{i} \in [n]$ or reports a failure, {\ie}, if $E_n = \varnothing$, then there is no element $i$ such that $i\in E_n$.

The procedure succeeds if it outputs an excellent element, namely if $\widehat{i}\in E_n$.  The total number of tests used by the procedure is $T_n = m_1 + m_2(Y_1)$.  Note that, we write $m_2(Y_1)$ instead of simply $m_2$ to signify that the number of second-round tests depends on the first-round outcomes.  Since the design is statistical, the relevant testing cost is the expected number of tests given by $\mathbb{E}_{\lambda,n}[T_n]$, where the expectation is taken over the random excellent set $E_n$ and over the induced test outcomes.

For a prescribed error tolerance $0 < \alpha < 1$, the admissible class consists of all two-round procedures satisfying the success constraint $\mathbb{P}_{\lambda,n}\{\widehat{i}\in E_n\}\ge 1-\alpha$.  The optimization problem studied in this paper is, therefore,
\begin{equation}
\begin{aligned}
&\min \mathbb{E}_{\lambda,n}[T_n] \\
&\text{subject to} \quad \mathbb{P}_{\lambda,n}\{\widehat{i}\in E_n\}\ge 1-\alpha .
\end{aligned}
\label{eq:main_optimization}
\end{equation}
over all two-round adaptive subset-testing procedures under the sparse model $X_i \stackrel{\text{indep.}}{\sim}\mathrm{Bernoulli}\left(\lambda/n\right)$, $i = 1, \ldots, n$.  The quantity of interest is the optimal expected number of tests, and is given by 
\begin{equation}
M_n^{(2)}(\lambda,\alpha) = \inf_{\Pi_n} \mathbb{E}_{\lambda,n}^{\Pi_n}[T_n],
\label{eq:expected_tests}
\end{equation}
where the infimum is taken over all admissible two-round procedures $\Pi_n$ satisfying $\mathbb{P}_{\lambda,n}^{\Pi_n}\{\widehat{i}\in E_n\}\ge 1-\alpha$.  The superscript $(2)$ in \eqref{eq:expected_tests} emphasizes the restriction to two rounds of searching.  The main result of our work is that $M_n^{(2)}(\lambda,\alpha)$ grows logarithmically in $n$ and is formalized in \thrmref{thrm:main_theorem}.

The problem formulation presented in this paper differs fundamentally from classical group testing described, for example, by \cite{Du2000}. In standard group testing, the usual goal is to recover the entire defective set, either exactly or with controlled error. The target object is the full set $E_n$. In our problem setup, the goal is strictly weaker: one only needs to find one excellent element. The identities of all other excellent elements are irrelevant. Moreover, the procedure is judged by success probability for producing a single excellent element, not by exact recovery of the whole support. This changes both the information requirement and the natural design criterion. A procedure may be successful even if it leaves most of $E_n$ unidentified.

The problem also differs from classical sparse group testing (see, for example, the work of \cite{Gandikota2019} for sparse group testing) in its cost criterion. We do not fix a deterministic number of tests and ask for exact support recovery. Instead, the second-round testing effort may depend on first-round outcomes, and the objective is the expected total number of tests. This makes the statistical two-round search problem closer to optimal sequential decision design than to standard support recovery. However, unlike fully sequential search, adaptivity is restricted to two rounds\textemdash preserving the combinatorial feature of the original problem posed by \cite{Katona2011}.

The novelty of our problem formulation is the combination of three features: rare independent excellence, two-round adaptivity, and a one-success objective. The sparse model $p_n=\lambda/n$ introduces a genuine statistical prior on the unknown excellent set; the two-round constraint retains the structural limitation of Katona's setting; and the success criterion asks only for one excellent element with probability at least $1-\alpha$. This produces an optimization problem that is neither deterministic two-round search nor classical group testing. It is a statistical two-round search problem in which the central object is the minimum expected number of noiseless subset tests required to find at least one excellent element with high probability.

The remainder of the article is organized as follows. \secref{sec:solution_search_problem} develops the main theorem and its proof.  \secref{sec:numerical_study} reports the numerical study.  Concluding remarks are presented in \secref{sec:conclusion} alongside some discussion on future research avenues on this problem.

\section{Two-round search in a statistical setting}\label{sec:solution_search_problem}
We solve the problem in the sparse regime $p_n = \lambda/n$,  $\lambda > 0$, where the excellence indicators $X_1, \dots, X_n$ are independent Bernoulli random variables with $\mathbb{P}(X_i = 1) = p_n$.  The set $E_n = \{i \in [n] : X_i = 1\}$ denotes the random set of excellent elements, and let $K_n = \lvert E_n \rvert $ denote the number of excellent elements. Then, we have 
\begin{equation}
K_n \sim \mathrm{Binomial}\left(n, \frac{\lambda}{n}\right).
\end{equation}
The subset test applied to $A \subseteq [n]$ is noiseless and returns $Y(A) = \mathbf{1}\left\{A \cap E_n \neq \varnothing \right\}$.  Thus a test is positive if and only if the tested subset contains at least one excellent element.

A summary of the two-round procedure is as follows. A procedure first chooses a collection of first-round test sets, observes their binary outcomes, then chooses a second-round collection of test sets. After the second round it outputs either an element $\widehat{i} \in [n]$ or, if $E_n = \varnothing$, then there is no element $i$ such that $i\in E_n$. The procedure succeeds if $\widehat{i} \in E_n$.  As described in \secref{subsec:problem_formulation}, the total number of tests is denoted by $T_n$. The goal is to minimize the expected number of tests given by $\mathbb{E}_{\lambda,n}[T_n]$, where the expectation is taken over the random excellent set $E_n$ and over the induced test outcomes, subject to $\mathbb{P}_{\lambda, n}\{\widehat{i} \in E_n\} \ge 1 - \alpha$.  We write $\gamma = 1 - \alpha$ for the required success probability. Thus the constraint is $\mathbb{P}_{\lambda, n}\{\widehat{i} \in E_n\} \ge \gamma$.  The optimal expected number of tests is $M_n^{(2)}(\lambda, \alpha) = \inf_{\Pi_n} \mathbb{E}_{\lambda, n}^{\Pi_n}[T_n]$, where the infimum is taken over all two-round procedures satisfying the success constraint.

The problem is not always feasible.  The infeasibility is due to the possible absence of an excellent element in the given population.  The probability that no element is excellent is
\begin{equation}
\mathbb{P}_{\lambda, n}(K_n = 0) = \left(1 - \frac{\lambda}{n}\right)^n \\
\implies \mathbb{P}_{\lambda, n}(K_n \ge 1) = 1 - \left(1 - \frac{\lambda}{n}\right)^n.
\end{equation}
No procedure\textemdash regardless of how many tests it uses\textemdash can succeed on the event $\{K_n = 0\}$.  Hence, every procedure satisfies
\begin{equation*}
\mathbb{P}_{\lambda, n}\{\widehat{i} \in E_n\} \le \mathbb{P}_{\lambda, n}(K_n \ge 1) = 1 - \left(1 - \frac{\lambda}{n}\right)^n.
\end{equation*}
Consequently, a necessary condition for feasibility is
\begin{equation}
\gamma \le 1 - \left(1 - \frac{\lambda}{n}\right)^n  \implies 1 - \alpha \le 1 - \left(1 - \frac{\lambda}{n}\right)^n \implies \alpha \ge \left(1 - \frac{\lambda}{n}\right)^n,
\label{eq:finite_n_feasible}
\end{equation}
which provides the finite-$n$ feasibility condition.  As $n \to \infty$,
\begin{equation}
\left(1 - \frac{\lambda}{n}\right)^n \longrightarrow e^{-\lambda}.
\end{equation}
Therefore, the asymptotic feasibility condition is $\alpha \ge e^{-\lambda}$.  If $\alpha < e^{-\lambda}$, then for all sufficiently large $n$ the success requirement is impossible.  This is a structural feature of the statistical formulation. The event that no excellent element exists is assigned a definite probability by the model, namely $\left(1-\lambda/n\right)^n$. Since no procedure can succeed on this event, it imposes an unavoidable upper bound on the attainable success probability.  Thus feasibility is governed not only by the testing design, but also by the prior probability that the population contains at least one excellent element.

Since $K_n \sim \mathrm{Binomial}\left(n, \lambda/n\right)$, we have the classical Poisson convergence $K_n \xrightarrow{d} K$,  where $K \sim \mathrm{Poisson}(\lambda)$.  For each fixed integer $k \ge 0$,
\begin{equation}
\mathbb{P}_{\lambda, n}(K_n = k) = \binom{n}{k} \left(\frac{\lambda}{n}\right)^k \left(1 - \frac{\lambda}{n}\right)^{n-k} \longrightarrow e^{-\lambda} \frac{\lambda^k}{k!}.
\end{equation}
In particular, $\mathbb{P}_{\lambda, n}(K_n = 0) \longrightarrow e^{-\lambda}$, and $\mathbb{P}_{\lambda, n}(K_n \ge 1) \longrightarrow 1 - e^{-\lambda}$.  Thus the maximum possible limiting success probability is $1 - e^{-\lambda}$.  The interesting regime is therefore $0 < \gamma < 1 - e^{-\lambda}$, or equivalently $e^{-\lambda} < \alpha < 1$.  In this regime, it is possible to find an excellent element with probability at least $\gamma$ for all sufficiently large $n$\textemdash a nontrivial problem because the excellent elements are sparse.

\subsection{Main result}\label{subsec:main_result}
We use standard asymptotic notation as $n \to \infty$. A quantity $a_n$ is $\mathcal{O}(\log n)$ if it is bounded above by a positive constant multiple of $\log n$ for all sufficiently large $n$. It is $\Omega(\log n)$ if it is bounded below by a positive constant multiple of $\log n$ for all sufficiently large $n$. It is $\Theta(\log n)$ if both bounds hold, so that $a_n$ grows on the logarithmic scale. Finally, $a_n = o(\log n)$ means that $a_n/\log n\to 0$, so $a_n$ grows strictly more slowly than $\log n$. The main finding of the paper is that the optimal expected number of tests grows logarithmically in $n$, and is made precise in the following theorem. 
\begin{theorem}
Fix $\lambda > 0$ and $\alpha \in (0, 1)$, and let $\gamma = 1 - \alpha$.  Assume $0 < \gamma < 1 - e^{-\lambda}$.  Then, there exist constants $0 < c_1(\lambda, \alpha) \le c_2(\lambda, \alpha) < \infty$ such that, for all sufficiently large $n$,
\begin{equation}
c_1(\lambda, \alpha) \log n \le M_n^{(2)}(\lambda, \alpha) \le c_2(\lambda, \alpha) \log n. 
\label{eq:matching_order}
\end{equation}
Equivalently,
\begin{equation}
M_n^{(2)}(\lambda, \alpha) = \Theta(\log n). 
\end{equation}
Thus in the sparse model $p_n = \lambda/n$, the minimum expected number of tests required by a two-round noiseless subset-testing procedure to find one excellent element with probability at least $1 - \alpha$ is logarithmic in the population size.  Proving an $\mathcal{O}(\log n)$ upper bound and an $\Omega(\log n)$ lower bound together establishes the sharp conclusion $\Theta(\log n)$ and rules out any $o(\log n)$ procedure.
\label{thrm:main_theorem}
\end{theorem}
The rest of this section proves \thrmref{thrm:main_theorem} in detail.  Before giving the proof, it is useful to explain why $\log n$ is the correct scale.  Suppose, for example, that there is exactly one excellent element. Conditional on $K_n = 1$, the excellent element is uniformly distributed over $[n]$. A successful procedure must identify its location. Since each test gives one binary answer, $t$ tests can produce at most $2^t$ different response patterns. To distinguish among $n$ possible locations, one expects at least $\log_2 n$ binary answers.  This is only a heuristic, because the procedure is not required to succeed for every possible excellent set. It only needs to succeed with probability at least $1 - \alpha$. Moreover, the number of excellent elements is random. Nevertheless, any fixed positive success probability forces the procedure to identify an element among $n$ possible labels on a non-negligible set of configurations. That requires logarithmically many bits of information.  The upper bound is also logarithmic. One first tests whether there is at least one excellent element. If the answer is positive, one applies a nonadaptive separating family of subset tests capable of identifying an excellent element whenever the number of excellent elements is not too large. In the sparse Poisson regime, the number of excellent elements is bounded by a fixed constant with probability as close as desired to $1 - e^{-\lambda}$. Hence, a fixed-size separating requirement suffices, and only $\mathcal{O}(\log n)$ tests are needed.  We now make this precise.

\subsection{A two-round construction attaining $\mathcal{O}(\log n)$ tests}\label{subsec:upper_bound}
We first prove the upper bound.  Fix $\gamma = 1 - \alpha$ with $0 < \gamma < 1 - e^{-\lambda}$.  Since $\mathbb{P}(K \ge 1) = 1 - e^{-\lambda}$, where $K \sim \mathrm{Poisson}(\lambda)$, there exists a positive integer $L$ such that $\mathbb{P}(1 \le K \le L) > \gamma$.  That is,
\begin{equation*}
\sum_{k=1}^L e^{-\lambda} \frac{\lambda^k}{k!} > \gamma.
\end{equation*}
The integer $L$ is a truncation level for the random number of excellent elements. It is chosen large enough so that the event of having between one and $L$ excellent elements carries at least the required success probability. Thus for the purpose of achieving success probability $1-\alpha$, the procedure only needs to guarantee success on the event $1\le K_n\le L$; the remaining event, where the number of excellent elements exceeds $L$, may be treated as part of the allowed failure probability.  Since $K_n \xrightarrow{d} K$, we also have, for all sufficiently large $n$,
\begin{equation}
\mathbb{P}_{\lambda, n}(1 \le K_n \le L) \ge \gamma. 
\label{eq:Kn_L}
\end{equation}
Thus it is enough to design a two-round procedure that succeeds whenever $1 \le K_n \le L$.  The constant $L$ depends on $\lambda$ and $\alpha$, but not on $n$.

We need a family of subsets that can identify excellent elements whenever the number of excellent elements is at most $L$.  We say a collection of subsets $\mathcal{F} = \{F_1, \dots, F_m\}$, $F_j \subseteq [n]$, is $L$-disjunct if, for every element $i \in [n]$ and every set
$S \subseteq [n] \setminus \{i\}$ with $\lvert S \rvert \le L$, there exists a test set $F_j \in \mathcal{F}$ such that $i \in F_j$ and $F_j \cap S = \varnothing$.  Equivalently, no column of the incidence matrix of $\mathcal{F}$ is contained in the Boolean union of any other $L$ columns.  This property is standard in nonadaptive group testing, but we use it here only for finding one excellent element.

Let $D \subseteq [n]$ be an unknown excellent set with $1 \le \lvert D \rvert \le L$.  Suppose we apply all tests in $\mathcal{F}$. The outcome vector is $Y_j = \mathbf{1}\{F_j \cap D \neq \varnothing\}$, $j = 1, \dots, m$.  An element $i$ is called compatible with the observed outcomes if every test containing $i$ is positive. That is,
\begin{equation*}
i \text{ is compatible} \quad \Longleftrightarrow \quad F_j \ni i \implies Y_j = 1 \quad \text{for every } j.
\end{equation*}
If $i \in D$, then every test containing $i$ must be positive, so every excellent element is compatible.  Now let $i \notin D$. Since $|D| \le L$ and $\mathcal{F}$ is $L$-disjunct, there exists $F_j \in \mathcal{F}$ such that $i \in F_j$ and $F_j \cap D = \varnothing$.  For this test, $Y_j = 0$.  Therefore, $i$ is not compatible.  Thus the compatible elements are exactly the excellent elements: $\{i : i \text{ is compatible}\} = D$.  Hence, if $1 \le |D| \le L$, the family $\mathcal{F}$ identifies the entire excellent set $D$. In particular, it identifies at least one excellent element.

For each fixed $L$, there exists an $L$-disjunct family on $[n]$ of size $m \le C_L \log n$, where $C_L$ is a constant depending only on $L$.  For completeness, we give the standard probabilistic proof.  Construct an $m \times n$ random binary matrix. Each entry is chosen independently as Bernoulli($q$), where $q = 1/(L+1)$.  The rows correspond to tests and the columns correspond to elements. A row contains element $i$ if the corresponding matrix entry is $1$.  Fix an element $i$ and a set $S \subseteq [n] \setminus \{i\}$, $\lvert S \rvert = s \le L$.  A row separates $i$ from $S$ if it contains $i$ and contains no element of $S$.  The probability that a given row separates $i$ from $S$ is $q(1-q)^s$.  Since $s \le L$, $q(1-q)^s \ge q(1-q)^L$.  With $q = 1/(L+1)$, we get
\begin{equation*}
q(1-q)^L = \frac{1}{L+1} \left(\frac{L}{L+1}\right)^L.
\end{equation*}
For notational convenience, let 
\begin{equation*}
\rho_L = \frac{1}{L+1} \left(\frac{L}{L+1}\right)^L.
\end{equation*}
Then, we have $\rho_L > 0$ for every fixed $L$.  The probability that none of the $m$ rows separates $i$ from $S$ is at most $(1 - \rho_L)^m \le e^{-\rho_L m}$.  The number of pairs $(i, S)$ with $\lvert S \rvert \le L$ is at most
\begin{equation}
n \sum_{s=0}^L \binom{n-1}{s} \le n \sum_{s=0}^L n^s \le (L+1)n^{L+1}. \label{eq:S_less_L}
\end{equation}
By the union bound, the probability that the random matrix fails to be $L$-disjunct is at most $(L+1)n^{L+1} e^{-\rho_L m}$.  Choose
\begin{equation*}
m \ge \frac{(L+2)\log n + \log(L+1)}{\rho_L}.
\end{equation*}
Then, $(L+1)n^{L+1} e^{-\rho_L m} \le n^{-1}$.  Therefore, for all sufficiently large $n$, there exists an $L-disjunct$ family with $m \le C_L \log n$, where one may take, for example, $C_L = (L + 3)/\rho_L$ for large $n$.  Since $L$ is fixed, $C_L$ is a constant independent of $n$.

We now define the two-round procedure.  In the first round, test the whole population: $A_{1, 1} = [n]$.  The first-round outcome is $Y([n]) = \mathbf{1}\{E_n \neq \varnothing\}$.  If $Y([n]) = 0$, then there is no excellent element, and the procedure declares failure.  If $Y([n]) = 1$,  then at least one excellent element exists, and the procedure proceeds to the second round.  Thus the first round uses exactly one test.

In the second round, choose an $L$-disjunct family $\mathcal{F} = \{F_1, \dots, F_m\}$ on $[n]$ with $m \le C_L \log n$.  If the first-round test is positive, apply all tests in $\mathcal{F}$.  After observing the outcomes, compute the compatible set
\begin{equation*}
\widehat{D} = \left\{ i \in [n] : F_j \ni i \implies Y(F_j) = 1 \text{ for all } j = 1, \dots, m \right\}.
\end{equation*}
If $\widehat{D} \neq \varnothing$, output any element of $\widehat{D}$.  If $\widehat{D} = \varnothing$, declare failure.

To verify the correctness, Suppose $1 \le K_n \le L$.  Then, $D = E_n$ is a nonempty set of size at most $L$.  Since $\mathcal{F}$ is $L$-disjunct, the compatibility argument above shows that $\widehat{D} = E_n$.  Therefore, the procedure outputs an element of $E_n$.
Hence, the procedure succeeds whenever $1 \le K_n \le L$.  Therefore, 
\begin{equation}
\mathbb{P}_{\lambda, n}\{\widehat{i} \in E_n\} \ge \mathbb{P}_{\lambda, n}(1 \le K_n \le L). 
\label{eq:correctness}
\end{equation}
By the choice of $L$, for all $n$ sufficiently large, $\mathbb{P}_{\lambda, n}(1 \le K_n \le L) \ge \gamma$.  Thus the procedure is admissible.

Let us now compute the expected number of tests.  The procedure always uses one first-round test.  It uses the $m$ second-round tests only if the first-round test is positive, that is, only if $K_n \ge 1$.  Therefore, the total number of tests is $T_n = 1 + m \mathbf{1}\{K_n \ge 1\}$.  Taking expectations, we get $\mathbb{E}_{\lambda, n}[T_n] = 1 + m \mathbb{P}_{\lambda, n}(K_n \ge 1)$.  But, $\mathbb{P}_{\lambda, n}(K_n \ge 1) \le 1$, so $\mathbb{E}_{\lambda, n}[T_n] \le 1 + m$.  Since $m \le C_L \log n$, we obtain $\mathbb{E}_{\lambda, n}[T_n] \le 1 + C_L \log n$.  Thus $M_n^{(2)}(\lambda, \alpha) \le 1 + C_L \log n$.  This proves the upper bound
\begin{equation}
M_n^{(2)}(\lambda, \alpha) = \mathcal{O}(\log n). 
\label{eq:upper_bound}
\end{equation}

\subsection{No admissible procedure can use $o(\log n)$ expected tests}\label{subsec:lower_bound}
We now prove that logarithmic growth is necessary.  The proof uses only elementary information-counting.  Let $\Pi_n$ be any two-round procedure satisfying $\mathbb{P}_{\lambda, n}^{\Pi_n}\{\widehat{i} \in E_n\} \ge \gamma$.  We show that $\mathbb{E}_{\lambda, n}^{\Pi_n}[T_n] \ge c \log n$ for some constant $c > 0$ depending only on $\lambda$ and $\alpha$.  Since $0 < \gamma < 1 - e^{-\lambda}$, we may choose an integer $L \ge 1$ such that $\mathbb{P}(1 \le K \le L) > \gamma/2$, $K \sim \mathrm{Poisson}(\lambda)$.  Equivalently, for sufficiently large $n$, $\mathbb{P}_{\lambda, n}(1 \le K_n \le L) > \gamma/2$.  But, for the lower bound we need a slightly different use of $L$.  Since $\mathbb{P}(K_n > L) \to \mathbb{P}(K > L)$, we may choose $L$ large enough so that, for all sufficiently large $n$, $\mathbb{P}_{\lambda, n}(K_n > L) < \gamma/2$.

Since the procedure has total success probability at least $\gamma$, we have
\begin{equation*}
\mathbb{P}_{\lambda, n}^{\Pi_n}\{\widehat{i} \in E_n, \ 1 \le K_n \le L\} \ge \gamma - \mathbb{P}_{\lambda, n}(K_n > L).
\end{equation*}
Hence, for all sufficiently large $n$, $\mathbb{P}_{\lambda, n}^{\Pi_n}\{\widehat{i} \in E_n, \ 1 \le K_n \le L\} \ge \gamma/2$.  Define the event $S_n = \{\widehat{i} \in E_n, \ 1 \le K_n \le L\}$.  Thus we have 
\begin{equation}
\mathbb{P}_{\lambda, n}^{\Pi_n}(S_n) \ge \frac{\gamma}{2}.
\label{eq:Pi_ge_gamma}
\end{equation}
In the following, we show that \eqref{eq:Pi_ge_gamma} forces the expected number of tests to be at least of order $\log n$.

Fix an integer $t \ge 0$.  Consider the event $\{S_n, \ T_n \le t\}$. On this event, the procedure succeeds, the number of excellent elements is between $1$ and $L$, and the procedure uses at most $t$ tests.  A binary testing procedure using at most $t$ tests can have at most $2^{t+1}$ terminal outcome patterns of length at most $t$.  This bound is elementary: a binary decision tree of depth at most $t$ has at most $1 + 2 + \dots + 2^t = 2^{t+1} - 1$ nodes, and therefore at most $2^{t+1}$ possible terminal outcomes.  At each terminal outcome, the procedure outputs either one element of $[n]$ or declares failure. If it outputs an element $i$, then for the output to be correct, the true excellent set must contain $i$.  Now fix $k$ with $1 \le k \le L$.

There are $\binom{n}{k}$ possible excellent sets of size $k$.  For a fixed output element $i$, the number of $k$-element excellent sets containing $i$ is $\binom{n-1}{k-1}$.  Therefore, the fraction of $k$-element sets for which a fixed output $i$ is correct is
\begin{equation*}
\frac{\binom{n-1}{k-1}}{\binom{n}{k}} = \frac{k}{n}.
\end{equation*}
Since there are at most $2^{t+1}$ terminal outcomes using at most $t$ tests, the fraction of $k$-element excellent sets on which the procedure can succeed using at most $t$ tests is at most
\begin{equation}
2^{t+1} \frac{k}{n} \le 2^{t+1} \frac{L}{n}.
\label{eq:t_tests}
\end{equation}
Thus conditional on $K_n = k$, using \eqref{eq:t_tests} gives us 
\begin{equation*}
\mathbb{P}_{\lambda, n}^{\Pi_n}\{\widehat{i} \in E_n, \ T_n \le t \mid K_n = k\} \le 2^{t+1} \frac{L}{n}.
\end{equation*}
Since this holds for every $1 \le k \le L$,
\begin{equation*}
\mathbb{P}_{\lambda, n}^{\Pi_n}\{S_n, \ T_n \le t\} \le 2^{t+1} \frac{L}{n}.
\end{equation*}
We now split the success event $S_n$ according to whether $T_n \le t$ or $T_n > t$:
\begin{equation}
\mathbb{P}_{\lambda,n}^{\Pi_n}(S_n) = \mathbb{P}_{\lambda,n}^{\Pi_n}(S_n, \ T_n \le t) + \mathbb{P}_{\lambda,n}^{\Pi_n}(S_n, \ T_n > t).
\label{eq:P_lambda}
\end{equation}
The first term in \eqref{eq:P_lambda} is bounded by the counting argument:
\begin{equation}
\mathbb{P}_{\lambda,n}^{\Pi_n}(S_n, \ T_n \le t) \le 2^{t+1} \frac{L}{n}.
\end{equation}
The second term in \eqref{eq:P_lambda} is bounded by $\mathbb{P}_{\lambda,n}^{\Pi_n}(T_n > t)$.  By Markov's inequality (see \cite[pp. 276]{Billingsley1995}), we have 
\begin{equation*}
\mathbb{P}_{\lambda,n}^{\Pi_n}(T_n > t) \le \frac{\mathbb{E}_{\lambda,n}^{\Pi_n}[T_n]}{t}.
\end{equation*}
Therefore, we can write
\begin{equation}
\mathbb{P}_{\lambda,n}^{\Pi_n}(S_n) \le 2^{t+1} \frac{L}{n} + \frac{\mathbb{E}_{\lambda,n}^{\Pi_n}[T_n]}{t}.
\label{eq:2t+1}
\end{equation}
But, from \eqref{eq:Pi_ge_gamma}, we already have 
\begin{equation*}
\mathbb{P}_{\lambda,n}^{\Pi_n}(S_n) \ge \frac{\gamma}{2}.
\end{equation*}
Hence, we get 
\begin{equation*}
\frac{\gamma}{2} \le 2^{t+1} \frac{L}{n} + \frac{\mathbb{E}_{\lambda,n}^{\Pi_n}[T_n]}{t} \implies \mathbb{E}_{\lambda,n}^{\Pi_n}[T_n] \ge t \left( \frac{\gamma}{2} - 2^{t+1} \frac{L}{n} \right).
\end{equation*}
Now choose $t = \left\lfloor \frac{1}{2} \log_2 n \right\rfloor$.  This choice of $t$ makes the ``few-test success'' term negligible.  The first term in \eqref{eq:2t+1}, namely $2^{t+1}\frac{L}{n}$, measures how much success is possible using at most $t$ binary tests. We want this term to go to zero as $n \to \infty$.  If we choose $t = \left\lfloor \frac{1}{2}\log_2 n\right\rfloor$, then $2^t \le 2^{(1/2)\log_2 n} = \sqrt{n}$.  Therefore, 
\begin{equation*}
2^{t+1}\frac{L}{n} \le 2\sqrt{n}\frac{L}{n} = \frac{2L}{\sqrt{n}} \to 0.
\end{equation*}
So, for large $n$, this term is smaller than, say, $\gamma/4$. Then
\begin{equation*}
\frac{\gamma}{2} \le \frac{\gamma}{4} + \frac{\mathbb{E}[T_n]}{t},
\end{equation*}
which implies $\frac{\mathbb{E}[T_n]}{t} \ge \frac{\gamma}{4}$.  Hence, $\mathbb{E}[T_n] \ge \frac{\gamma}{4}t$.  Since $t=\left\lfloor \frac{1}{2}\log_2 n\right\rfloor$, this gives $\mathbb{E}[T_n] \ge c\log n$ for some constant $c > 0$.  So, $c = 1/2$ is a convenient choice that makes $2^t$ grow like $\sqrt{n}$, which is much smaller than $n$. More generally, we could choose $t = \lfloor c\log_2 n\rfloor$ for any fixed $0 < c < 1$.  Then, $2^t \le n^c$, and $2^{t+1}\frac{L}{n} \le 2L n^{c-1} \to 0$.  The specific choice $c = 1/2$ is not essential; it is for simplicity.  Then, we have 
\begin{equation*}
2^{t+1} \frac{L}{n} \le 2L \frac{\sqrt{n}}{n} = \frac{2L}{\sqrt{n}}.
\end{equation*}
Therefore, $2^{t+1} \frac{L}{n} \longrightarrow 0$.  For sufficiently large $n$, $2^{t+1} \frac{L}{n} \le \frac{\gamma}{4}$.  Thus we get 
\begin{equation*}
\mathbb{E}_{\lambda,n}^{\Pi_n}[T_n] \ge t \left( \frac{\gamma}{2} - \frac{\gamma}{4} \right) = \frac{\gamma}{4}t.
\end{equation*}
Since $t = \left\lfloor \frac{1}{2} \log_2 n \right\rfloor$, we obtain $\mathbb{E}_{\lambda,n}^{\Pi_n}[T_n] \ge \frac{\gamma}{8} \log_2 n$ for all sufficiently large $n$, up to an immaterial additive constant.  Equivalently, using natural logarithms,
\begin{equation*}
\mathbb{E}_{\lambda,n}^{\Pi_n}[T_n] \ge \frac{\gamma}{8 \log 2} \log n + O(1).
\end{equation*}
Since $\Pi_n$ was arbitrary, this implies $M_n^{(2)}(\lambda, \alpha) \ge c_1(\lambda, \alpha) \log n$ for some constant $c_1(\lambda, \alpha) > 0$.  This proves the lower bound
\begin{equation}
M_n^{(2)}(\lambda, \alpha) = \Omega(\log n).
\end{equation}
Combining the lower and upper bounds gives
\begin{equation}
M_n^{(2)}(\lambda, \alpha) = \Theta(\log n).
\label{eq:growth_order}
\end{equation}
This completes the proof of \thrmref{thrm:main_theorem}, which identifies the sharp growth order shown in \eqref{eq:growth_order}.

In the sparse model $p_n = \lambda/n$, the expected number of excellent elements is $\mathbb{E}[K_n] = \lambda$.  Thus the excellent set has constant expected size, even as $n$ grows. The search problem is, therefore, essentially a sparse-location problem. To find one excellent element, the procedure must acquire enough binary information to isolate an index among $n$ possible labels on a non-negligible subset of outcomes. This requires order $\log n$ tests.  The two-round constraint does not increase the order beyond $\log n$. A first-round global test detects whether at least one excellent element exists. Conditional on existence, a second-round disjunct family can locate an excellent element whenever the number of excellent elements is bounded by a fixed integer $L$. Since $K_n$ converges to a Poisson random variable, choosing $L$ large enough captures any prescribed success probability strictly below the maximum possible value $1 - e^{-\lambda}$. Thus for fixed $\lambda$ and fixed feasible $\alpha$ with $e^{-\lambda} < \alpha < 1$, the solution is $M_n^{(2)}(\lambda, \alpha) = \Theta(\log n)$.  The exact multiplicative constant is not determined by this argument.  Determining the exact asymptotic constant would require a finer optimization over two-round testing designs and is a separate refinement problem.

\subsection{Discussion}\label{subsec:discussion}
Firstly, there are three distinct regimes of operation for the statistical two-round search procedure. 
\begin{enumerate}[(a)]
\item Infeasible regime: If $\gamma > 1 - \left( 1 - \frac{\lambda}{n} \right)^n$, then no procedure can satisfy the success constraint.  Equivalently, if $\alpha < \left( 1 - \frac{\lambda}{n} \right)^n$, then $M_n^{(2)}(\lambda, \alpha) = +\infty$.  Asymptotically, if $\alpha < e^{-\lambda}$, then the problem is infeasible for all sufficiently large $n$.

\item Nontrivial feasible regime: If $0 < \gamma < 1 - e^{-\lambda}$, or equivalently $e^{-\lambda} < \alpha < 1$, then $M_n^{(2)}(\lambda, \alpha) = \Theta(\log n)$.  This is the main regime of this paper.

\item Trivial low-success regime:  If $\gamma = 0$, or equivalently $\alpha = 1$, then no success is required. The empty procedure, using no tests and declaring failure, is admissible. Therefore, $M_n^{(2)}(\lambda, 1) = 0$.
\end{enumerate}
The construction used an $L$-disjunct family, which is a familiar object from group testing. However, the present problem is not a standard group-testing problem.  The distinction is structural.  In classical group testing, the goal is usually to recover the entire defective set $E_n$. A procedure is successful only if it identifies every defective and every nondefective element correctly, or if it satisfies some approximate support-recovery criterion.  Here the goal is much weaker: find one element of $E_n$.  The procedure does not need to estimate $E_n$. It does not need to know how many excellent elements there are. It does not need to classify all non-excellent elements. It only needs to output one index $\widehat{i}$ satisfying $\widehat{i} \in E_n$.  This changes the statistical objective. The success probability is $\mathbb{P}_{\lambda,n}(\widehat{i} \in E_n)$, not $\mathbb{P}_{\lambda,n}(\widehat{E}_n = E_n)$.  The cost criterion is also different. We minimize $\mathbb{E}_{\lambda,n}[T_n]$, allowing the second-round testing burden to depend on the first-round outcome. In particular, if the first-round test of the whole population is negative, the procedure stops immediately. This is natural in the present problem because, on the event $E_n = \varnothing$, success is impossible.  In standard group testing, especially in exact recovery formulations, the no-defective case is just another support pattern to be identified. In the present search problem, it is an unavoidable failure event, and it directly determines the feasibility boundary $\alpha \ge \left( 1 - \frac{\lambda}{n} \right)^n$.  Thus group-testing designs are useful as tools, but the optimization problem is different. We use disjunct families only to guarantee that, with high probability under the sparse model, at least one excellent element can be certified. The paper's objective is not support recovery; it is high-probability one-success search under a two-round expected-cost criterion.

From \eqref{eq:matching_order}, we see that both bounds are on the same scale, namely $\log n$, and they are called matching-order bounds.  A matching-order upper bound is achieved by: (i) testing the whole population in the first round;(ii) if the test is positive, applying an $L$-disjunct family in the second round; and (iii) choosing $L$ so that $\mathbb{P}(1 \le \operatorname{Poisson}(\lambda) \le L) > 1 - \alpha$.  A matching-order lower bound follows from a counting argument: with fewer than order $\log n$ binary test outcomes, a procedure cannot identify an element among $n$ possible labels with fixed positive probability on the event that the number of excellent elements is bounded.  Thus the two-round statistical version of Katona's search problem has a clean logarithmic solution in the rare-excellence regime.

\section{Numerical studies}\label{sec:numerical_study}
The preceding section establishes the order result $M_n^{(2)}(\lambda, \alpha) = \Theta(\log n)$ in the nontrivial feasible regime $e^{-\lambda} < \alpha < 1$.  The purpose of this section is to illustrate three concrete features of the theory:  (i) the feasibility boundary imposed by the event of zero excellent elements; (ii) the Poisson truncation level $L$ needed to guarantee the desired success probability; and (iii)the logarithmic growth of the constructive upper bound.  Recall that $K_n = |E_n| \sim \mathrm{Binomial}\left(n, \lambda/n\right)$, and $K_n \xrightarrow{d} K$, where $K \sim \mathrm{Poisson}(\lambda)$.  Therefore, the limiting probability that at least one excellent element exists is $\mathbb{P}(K \ge 1) = 1 - e^{-\lambda}$.  No procedure can succeed when $K = 0$. Hence, the limiting success probability cannot exceed $1 - e^{-\lambda}$.  This gives the feasibility condition $1 - \alpha \le 1 - e^{-\lambda}$, or equivalently, $\alpha \ge e^{-\lambda}$.

\subsection{Feasibility boundary}\label{subsec:feasible_boundary}
\tabref{tab:feasible_boundary} gives the limiting probability of no excellent element and the corresponding maximum possible success probability.
\begin{table}[ht]
\centering
\begin{tabular}{|c|c|c|}
\hline
$\lambda$ & $e^{-\lambda}$ & $1 - e^{-\lambda}$ \\
\hline
0.5 & 0.6065 & 0.3935 \\
1.0 & 0.3679 & 0.6321 \\
1.5 & 0.2231 & 0.7769 \\
2.0 & 0.1353 & 0.8647 \\
3.0 & 0.0498 & 0.9502 \\
5.0 & 0.0067 & 0.9933 \\
8.0 & 0.0003 & 0.9997 \\
\hline
\end{tabular}
\caption{Limiting feasibility boundary for the statistical  two-round search problem.}
\label{tab:feasible_boundary}
\end{table}
The table shows an important qualitative feature of the problem. When $\lambda$ is small, high success probability is impossible, regardless of how many tests are used. For example, if $\lambda = 1$, then $\mathbb{P}(K = 0) = e^{-1} \approx 0.3679$.  Thus the maximum possible success probability is only $1 - e^{-1} \approx 0.6321$.  Consequently, a target success probability such as $0.90$ or $0.95$ is infeasible when $\lambda = 1$. This is not a limitation of the proposed two-round design; it is an intrinsic limitation of the statistical model.  On the other hand, when $\lambda$ is moderately large, the probability of having no excellent element becomes small. For $\lambda = 5$, the maximum possible success probability is approximately $0.9933$, so targets such as $0.90$ and $0.95$ are feasible.

\subsection{Poisson truncation level needed for the construction}\label{subsec:Poisson_truncation}
\begin{table}[htbp!]
\centering
\begin{tabular}{|c|c|c|c|}
\hline
$\lambda$ & $1-\alpha$ & smallest $L$ & $\mathbb{P}(1 \le K \le L)$ \\
\hline
1 & 0.50 & 2 & 0.5518 \\
1 & 0.80 & infeasible & -- \\
1 & 0.90 & infeasible & -- \\
1 & 0.95 & infeasible & -- \\
\hline
2 & 0.50 & 2 & 0.5413 \\
2 & 0.80 & 4 & 0.8120 \\
2 & 0.90 & infeasible & -- \\
2 & 0.95 & infeasible & -- \\
\hline
3 & 0.50 & 3 & 0.5974 \\
3 & 0.80 & 5 & 0.8663 \\
3 & 0.90 & 6 & 0.9167 \\
3 & 0.95 & 11 & 0.9501 \\
\hline
5 & 0.50 & 5 & 0.6092 \\
5 & 0.80 & 7 & 0.8599 \\
5 & 0.90 & 8 & 0.9252 \\
5 & 0.95 & 9 & 0.9614 \\
\hline
\end{tabular}
\caption{Smallest Poisson truncation level $L$ satisfying $\mathbb{P}(1 \le K \le L) \ge 1 - \alpha$.}
\label{tab: Poisson_truncation}
\end{table}
The proof of the upper bound uses an integer $L$ such that $\mathbb{P}(1 \le K \le L) \ge 1 - \alpha$, where $K \sim \mathrm{Poisson}(\lambda)$.  Equivalently, $L$ is chosen so that $e^{-\lambda} \sum_{k=1}^L \frac{\lambda^k}{k!} \ge 1 - \alpha$.  The integer $L$ determines how many excellent elements the second-round separating design must be able to handle. \tabref{tab: Poisson_truncation} gives the smallest such $L$ for selected values of $\lambda$ and target success probability $1 - \alpha$.  This table gives a useful operational interpretation of the \thrmref{thrm:main_theorem}. The construction does not need to solve the search problem for all possible values of $K$. It only needs to succeed on the event $1 \le K \le L$, whose probability is at least the target success probability. Since $K$ has a Poisson distribution with fixed mean $\lambda$, the required $L$ remains fixed as $n$ grows.  For example, when $\lambda = 5$ and the desired success probability is $0.95$, it is enough to guarantee success whenever $1 \le K \le 9$.  The remaining event $\{K \ge 10\}$ has small enough probability that failure there is allowed under the prescribed risk level. This is precisely why the sparse Poisson regime leads to a logarithmic number of tests rather than powers of the population size $n$. 

\subsection{Constructive logarithmic upper bound}\label{subsec:log_upper_bound}
\begin{table}[ht]
\centering
\begin{tabular}{|c|c|c|}
\hline
$L$ & $\rho_L$ & $C_L = (L+3)/\rho_L$ \\
\hline
1 & 0.2500 & 16.00 \\
2 & 0.1481 & 33.75 \\
3 & 0.1055 & 56.89 \\
4 & 0.0819 & 85.45 \\
5 & 0.0670 & 119.44 \\
6 & 0.0567 & 158.86 \\
7 & 0.0491 & 203.72 \\
8 & 0.0433 & 254.01 \\
9 & 0.0387 & 309.74 \\
10 & 0.0350 & 370.91 \\
11 & 0.0320 & 437.51 \\
\hline
\end{tabular}
\caption{Constants arising from the elementary random-construction proof of an $L$-disjunct family.}
\label{tab:log_upper_bound}
\end{table}
For the analytical proof, we used an $L$-disjunct family. A standard probabilistic construction gives an $L$-disjunct family of size at most $m_L(n) = \lceil C_L \log n \rceil$, where one possible proof constant is $C_L = \frac{L + 3}{\rho_L}$, with 
\begin{equation*}
\rho_L = \frac{1}{L + 1} \left( \frac{L}{L + 1} \right)^L.
\end{equation*}
The resulting two-round construction uses one first-round test of the full population. If this test is positive, it applies the $m_L(n)$ second-round tests. Thus $T_n = 1 + m_L(n) \mathbf{1}_{\{K_n \ge 1\}}$.  Hence, $\mathbb{E}[T_n] = 1 + m_L(n) \mathbb{P}(K_n \ge 1)$.  For large $n$, $\mathbb{P}(K_n \ge 1) \approx 1 - e^{-\lambda}$.  Therefore, the asymptotic constructive bound is 
\begin{equation*}
\mathbb{E}[T_n] \lesssim 1 + (1 - e^{-\lambda}) C_L \log n.
\end{equation*}
\tabref{tab:log_upper_bound} displays the proof constant $C_L$ for selected values of $L$.  The constants in \tabref{tab:log_upper_bound} are conservative. They come from a union-bound proof, not from an optimized design.  Therefore, they should not be interpreted as sharp constants for the true optimum $M_n^{(2)}(\lambda, \alpha)$. Their purpose is to make the upper-bound proof explicit and computable.  The important point is that, for each fixed $L$, the second-round design has size proportional to $\log n$.  Thus the dependence on the population size is logarithmic.

\subsection{Logarithmic scaling}\label{subsec:log_scaling}
\begin{table}[ht]
\centering
\begin{tabular}{|c|c|c|c|c|c|}
\hline
$\lambda$ & $1-\alpha$ & $L$ & $n$ & $m_L(n)$ & $\mathbb{E}[T_n]$ upper bound \\
\hline
1 & 0.50 & 2 & $10^3$ & 234 & 148.96 \\
1 & 0.50 & 2 & $10^6$ & 467 & 296.20 \\
1 & 0.50 & 2 & $10^9$ & 700 & 443.48 \\
\hline
2 & 0.80 & 4 & $10^3$ & 591 & 512.18 \\
2 & 0.80 & 4 & $10^6$ & 1{,}181 & 1{,}022.17 \\
2 & 0.80 & 4 & $10^9$ & 1{,}771 & 1{,}532.32 \\
\hline
3 & 0.90 & 6 & $10^3$ & 1{,}098 & 1{,}044.58 \\
3 & 0.90 & 6 & $10^6$ & 2{,}195 & 2{,}086.72 \\
3 & 0.90 & 6 & $10^9$ & 3{,}293 & 3{,}130.05 \\
\hline
5 & 0.95 & 9 & $10^3$ & 2{,}140 & 2{,}126.76 \\
5 & 0.95 & 9 & $10^6$ & 4{,}280 & 4{,}252.16 \\
5 & 0.95 & 9 & $10^9$ & 6{,}419 & 6{,}376.75 \\
\hline
\end{tabular}
\caption{Constructive upper bound on the expected number of tests.}
\label{tab:log_scaling}
\end{table}
\tabref{tab:log_scaling} gives the constructive upper bound for several representative choices of $(\lambda, 1 - \alpha)$ and $n$.  For each row, $L$ is the smallest integer satisfying $\mathbb{P}(1 \le K \le L) \ge 1 - \alpha$, where $K \sim \operatorname{Poisson}(\lambda)$. This means that the second-round design is required to succeed whenever the number of excellent elements is between one and $L$. The number of second-round tests in the constructive design is $m_L(n) = \lceil C_L \log n \rceil$. Since the first round consists of one test of the whole population, this one test is always used. The second-round tests are used only if the first-round test is positive, which occurs precisely when $K_n \ge 1$.  Since $K_n \sim \operatorname{Binomial}(n, \lambda/n)$, this probability is $1 - (1 - \lambda/n)^n$. Therefore, the expected number of tests for this construction is 
\begin{equation*}
\mathbb{E}[T_n] = 1 + \left\{1 - \left(1 - \frac{\lambda}{n}\right)^n\right\} m_L(n).
\end{equation*}
The important feature of \tabref{tab:log_scaling} is the scaling in $n$. Increasing $n$ from $10^3$ to $10^6$ multiplies the population size by $1,000$, but it approximately doubles $\log n$. Accordingly, the number of tests approximately doubles.  Increasing $n$ again from $10^6$ to $10^9$ approximately doubles neither the population nor the expected number of tests used by the procedure; it merely adds another constant multiple of $\log(10^3)$.  For example, for $(\lambda, 1 - \alpha) = (1, 0.50)$, the constructive second-round test size increases from $234$ to $467$ to $700$ as $n$ moves from $10^3$ to $10^6$ to $10^9$, respectively. This is exactly the logarithmic pattern predicted by \thrmref{thrm:main_theorem}.  The larger values in the lower rows are not caused by $n$ alone. They are primarily caused by the stronger success requirement, which forces a larger truncation level $L$. For instance, achieving success probability $0.95$ when $\lambda = 5$ requires covering the event $1 \le K \le 9$.  The corresponding disjunctness requirement is much stronger than for $L = 2$, and therefore, the proof constant $C_L$ is larger. This illustrates that the dependence on $n$ is logarithmic, but the constant multiplying $\log n$ depends on how much of the Poisson mass one wishes to cover.

These numerical results support the analytical conclusions in three ways.  First, the feasibility boundary is substantial. The condition $1 - \alpha \le 1 - e^{-\lambda}$ is not a technical artifact. For small $\lambda$, the event of ``no excellent elements'' has large probability, so high-probability success is impossible.  Second, the Poisson truncation level $L$ provides a simple bridge between the statistical model and the combinatorial design. The statistical requirement determines how much of the Poisson distribution must be covered. Once $L$ is chosen, the second-round testing problem becomes a finite-$L$ separating problem.  Third, the dependence on $n$ is mild. The number of tests grows like $\log n$, not like a power of $n$. This is the central quantitative conclusion of the theory.  The numerical values in \tabref{tab:log_upper_bound} and \tabref{tab:log_scaling} should not be read as optimized practical test counts. They come from a deliberately elementary random-construction argument. Sharper explicit constructions or optimized two-round designs may substantially improve the constants. However, such refinements would not change the result that $M_n^{(2)}(\lambda, \alpha) = \Theta(\log n)$ throughout the nontrivial feasible regime.

\section{Conclusion and future directions}\label{sec:conclusion}
This paper introduced a statistical two-round search problem motivated by Katona's two-round framework for finding an excellent element. The essential modification is probabilistic: each element is independently excellent with probability $\lambda/n$, subset tests are noiseless, and the objective is to find one excellent element with probability at least $1-\alpha$ while minimizing the expected number of tests. This formulation separates the problem from classical group testing. The goal is not support recovery, nor even approximate support recovery, but one-success identification under a two-round expected-cost criterion.  A central feature of the model is that success is intrinsically limited by the possibility that no excellent element exists. Thus the feasibility boundary is not an artifact of the proof, but a structural property of the problem. In the sparse regime, the number of excellent elements converges to a Poisson random variable, and the maximum possible limiting success probability is $1-e^{-\lambda}$. Within the nontrivial feasible regime, the optimal expected number of tests grows logarithmically in the population size. The proof combines a constructive upper bound, based on a first-round global existence test followed by a second-round separating design, with an information-counting lower bound showing that fewer than logarithmically many binary outcomes cannot identify an excellent element with fixed positive probability.

The present work leaves several natural problems open.  A few are listed below. 
\begin{enumerate}[(i)]
\item Sharp asymptotic constants:  \thrmref{thrm:main_theorem} establishes the order $M_n^{(2)}(\lambda,\alpha)=\Theta(\log n)$, but it does not identify the exact asymptotic constant. It is useful to determine whether the limit
\begin{equation*}
\lim_{n\to\infty} \frac{M_n^{(2)}(\lambda,\alpha)}{\log n}
\end{equation*}
exists, and if so, to characterize it as a function of $\lambda$ and $\alpha$.  The problem can be formulated in terms of information per test under the sparse prior. Since the objective is to find one excellent element rather than the full excellent set, the relevant entropy is not simply the entropy of the random support $E_n$. One needs an information measure adapted to partial identification. This suggests studying quantities such as Shannon's information entropy $I(\widehat{i};E_n)$ treated by \cite[Chapter 2.2]{Yeung2002} or, more directly, the maximum posterior probability that a candidate index is excellent after a given number of tests. Fano-type arguments explained by \cite[Chapter 2.8]{Yeung2002} may provide sharper lower bounds, while random coding methods shown by \cite[Chapter 8.4]{Yeung2002} may provide improved upper bounds.

\item Optimal two-round designs beyond disjunct families:  The second-round design used in this paper is deliberately strong. An $L$-disjunct family identifies the entire excellent set whenever its size is at most $L$.  But, the statistical search problem only requires one excellent element. Therefore, full disjunctness is more than needed.  This raises the following problem: construct and analyze smaller two-round families that certify at least one excellent element without necessarily identifying all excellent elements. Such designs may have substantially better constants than $L$-disjunct families.  One possible direction is to define a weaker combinatorial object. For example, a family of tests could be called one-finding for sets of size at most $L$ if, for every nonempty excellent set $D$ with $|D| \le L$, the test outcomes force at least one identifiable member of $D$, even if they do not identify all of $D$. This notion lies between arbitrary separating families and fully disjunct matrices. Determining the minimum size of such one-finding families would give a sharper upper bound for the statistical problem.  

\item Noisy subset tests:  The present paper assumes noiseless subset tests. In many applications this assumption is idealized. A natural extension is to allow false positives and false negatives. For example, one may assume
\begin{eqnarray*}
\mathbb{P}(Y(A)=1 \mid A \cap E_n \neq \varnothing) &=& 1-\beta_-, \\
\mathbb{P}(Y(A)=1 \mid A \cap E_n = \varnothing) &=& \beta_+,
\end{eqnarray*}
where $\beta_-$ and $\beta_+$ are error probabilities.  Noisy tests create two new issues. First, the global existence test is no longer definitive. A negative first-round test does not prove that no excellent element exists, and a positive test may be a false alarm. Second, the second-round decoding problem becomes probabilistic rather than logical.  A likely approach is to replace deterministic separating families with error-correcting test designs. Repetition, majority decoding, and likelihood-based decoding could be incorporated. The main question is whether the optimal expected number of tests remains logarithmic in $n$ and how the constant depends on the noise parameters. One expects logarithmic scaling to persist for fixed noise levels bounded away from complete randomness, but the proof would require information-theoretic lower bounds and robust coding constructions.

\item More than two rounds:  Katona's original motivation emphasizes the restriction to two rounds. Nevertheless, it is natural to compare two-round performance with fully sequential or multi-round search.  With more rounds, one can use adaptive bisection-type strategies once a positive subset has been located. For example, after determining that an excellent element exists, a sequential binary splitting strategy can isolate one excellent element in about $\log_2 n$ additional tests. The question is whether additional adaptivity improves only constants or changes the expected-test structure in some regimes.  A systematic extension would define $M_n^{(r)}(\lambda,\alpha)$ as the optimal expected number of tests using at most $r$ rounds. Natural questions include whether $M_n^{(r)}(\lambda,\alpha)=\Theta(\log n)$ for every fixed $r \ge 2$, how the constants depend on $r$, and whether the limit as $r \to \infty$ matches the fully sequential optimum.  This direction would clarify precisely what price is paid for the two-round restriction.
\end{enumerate}

\vspace{-0.25cm}
\section*{Disclosure of interest}
The authors report there are no competing interests to declare.

\bibliography{/Users/kgnagananda/Documents/Work/collaborations/pdx/research/references/research_pdx.bib}

\end{document}